\documentclass[]{aa}
\usepackage{graphicx}
\begin{document}
\title{Moments of inertia for neutron and strange stars: limits derived 
for the Crab pulsar}

\author{M. Bejger \and P. Haensel}
\institute{
N. Copernicus Astronomical Center, Polish Academy of
Sciences, Bartycka 18, PL-00-716 Warszawa, Poland\\
e-mail: {\tt  bejger@camk.edu.pl, haensel@camk.edu.pl}
}
\offprints{M. Bejger}
\date{Received 28/03/02 /Accepted 27/08/02}
\abstract{
Recent estimates of the properties of the Crab nebula are used to 
derive constraints on the moment of inertia, mass and radius of the 
pulsar. To this purpose, we employ an approximate formula combining 
these three parameters.
Our ``empirical formula'' $I\simeq a(x) M R^2$, where $x=(M/M_\odot) ({\rm km}/R)$,
is based on numerical results obtained for thirty theoretical
equations of state of dense matter.
The functions $a(x)$ for neutron stars
and strange stars are qualitatively different. For neutron stars
$a_{\rm NS}(x)=x/(0.1+2x)$ for $x\le 0.1$ (valid
for $M>0.2~{\rm M}_\odot$)
and $a_{\rm NS}(x)={2\over 9}(1+5x)$ for $x>0.1$.
For strange
stars  $a_{\rm SS}(x)={2\over 5}(1+x)$ (not valid for strange stars
with crust and $M<0.1~M_\odot$).
We obtain also an approximate expression for the maximum moment of
inertia $I_{\rm max,45}\simeq
(-0.37 + 7.12\cdot x_{\rm max})
(M_{\rm max}/M_\odot)(R_{M_{\rm max}}/{10~{\rm km}})^2$, 
where $I_{\rm 45} = I/10^{45}~{\rm g\cdot cm^2}$, valid for
both neutron stars and strange stars.
Applying our formulae to the evaluated values of $I_{\rm Crab}$, 
we derive constraints on the mass and radius of the pulsar.
{ A very conservative evaluation of the expanding nebula mass, 
$M_{\rm neb}=2~M_{\odot}$, yields $M_{\rm Crab}>1.2~M_{\odot}$ and 
$R_{\rm Crab}=10-14~{\rm km}$. Setting the most recent evaluation
(``central value'') $M_{\rm neb}=4.6~M_{\odot}$
rules out most of the existing equations
of state, leaving only the stiffest ones: $M_{\rm Crab}>1.9~M_{\odot}$,
$R_{\rm Crab}=14-15~{\rm km}$.
}
}
\titlerunning{Neutron and strange stars moment of inertia}
\authorrunning{M. Bejger and P. Haensel}
\maketitle
\section{Introduction}
The moment of inertia of neutron stars plays a crucial role
in the models of radio pulsars. In the standard case of
rigid rotation the total energy expenditure per unit time
$\dot{E}_{\rm tot}$ is related to the measured pulsar angular
frequency $\Omega$ and its time derivative $\dot{\Omega}$ by
$\dot{E}_{\rm tot}=-I\Omega{\dot{\Omega}}$. However, even in the
simplest  case of slow rotation
($\Omega^2\ll (c/R)^2GM/Rc^2$, $R\Omega\ll c$) characteristic
of  observed radio pulsars, the relation of
$I$ to the matter distribution within the star
is complicated by the general relativistic effects such
as the dragging of the local inertial frames.
 Among all global neutron star parameters the moment of inertia is
the most sensitive to the dense matter equation of state (EOS).
The ``theoretical maximum mass'' of neutron stars increases by a factor
of two when going from the softest to the stiffest EOS.
The ``theoretical maximum moment of inertia'' however
increases then by a factor of seven (Haensel 1990).
In view of this particular sensitivity of $I$ to the
largely unknown equation of state (EOS) of dense matter at
supra-nuclear density, it is of interest to look for
observational evaluations $I$, which could then be used
to constrain theoretical models.

In the present paper we use recent evaluations of the parameters
of the expanding Crab nebula, in order to derive constraints on
the Crab-pulsar neutron star. In Sect. 2
 we reconsider the energetics of the Crab nebula, and using
 expanding-shell model of the  nebula we derive a constraint on
the moment of inertia of the Crab pulsar.
 In Sect. 3 we derive approximate formulae for $I$
 of neutron stars and strange
stars, based on a statistical analysis of a
 numerical data sample obtained
for thirty theoretical  EOSs of dense matter.
These formulae are then used in Sect. 4
to derive constraints in the $M-R$ plane,
implied by the value of $I_{\rm Crab}$,
 deduced in Sect.\ 2.
 Appendix presents results of an analysis of the
 correlation between the maximum value of $I$ and the mass and radius  of
the configuration with maximum allowable mass.
\section{Energetics of the Crab nebula and constraints on $I_{\rm Crab}$}
The present values of $\Omega$ and $\dot\Omega$ for the Crab pulsar are $188.119~{\rm
s^{-1}}$ and $-2.366\cdot 10^{-13}~{\rm s^{-2}}$, respectively (Lyne~\&~Graham-Smith
1998). Therefore, the present rate of the rotational energy loss is
\begin{equation}
I\Omega \dot{\Omega}=\dot{E}_{\rm rot}= 4.45\cdot 10^{38}~I_{45}~{\rm erg~s^{-1}}~.
\label{dErotdt.Crab}
\end{equation}
It is commonly accepted that $\dot{E}_{\rm rot}$ has to account for the luminosity of
the pulsar and the nebula, as well as for the accelerated expansion of the nebula, so
that
\begin{equation}
\dot{E}_{\rm rot} > \dot{E}_{\rm rad} +\dot{E}_{\rm exp}~. \label{E.balance}
\end{equation}
Although the nebula has a complicated shape we will assume for simplicity 
that it is spherically symmetric with a radius $R_{\rm neb}\simeq 1.25~$pc 
(see e.g. Douvion et al. 2001, Petersen 1998). The total power of
radiation in the X, UV, and optical domains is given by Petersen (1998):
\begin{equation}
\dot{E}_{\rm rad}\simeq 1.5\cdot 10^{38}~\left({d\over{\rm 2~kpc}}\right)^2~{\rm
erg~s^{-1}}~. \label{E.rad}
\end{equation}
{ 
Nugent (1998) calculated the velocities of expansion of 50 bright 
filaments by comparing their positions on four high-resolution 
photographs taken in 1939, 1960, 1976 and 1992. After projecting 
the motion of filaments backward in time, he obtained convergence
in AD $1130 \pm 16$ yr, in accordance with Trimble (1968). The calculated
moment of the supernova event is by $\Delta=76\ {\rm yr}$ later 
than the date recorded by the Chinese astronomers, AD 1054. Therefore, 
the actual expansion had a non-zero acceleration. Assuming a constant
value of the acceleration, $\dot{v}$, and putting the 
present mean expansion velocity, determined from the measurements of 
the spectra, 
$v\sim 1.5\cdot 10^8~{\rm cm~s^{-1}}$ (e.g. Davidson \& Fesen 1985, 
Bietenholtz et al. 1991, Sollerman et al. 2000), we obtain

\begin{equation}
\label{acceleration}
\dot{v} = \frac{2v\Delta}{T^2}\simeq 0.82
\cdot 10^{-3}~
{\rm cm~s^{-2}}.
\end{equation}

}

The Crab nebula expands in the interstellar, mostly hydrogenic, gas. Let the 
number density of the hydrogen atoms be $n_{\rm H}$. 
The expanding shell sweeps the
interstellar medium, which becomes a part of the shell after being accelerated to the
shell velocity. Consequently, the mass of the nebula increases at the rate
\begin{equation}
\dot{M}_{\rm neb}=4\pi R_{\rm neb}^2 n_{\rm H} m_{\rm H} v~, \label{Mneb.rate}
\end{equation}
where $m_{\rm H}$ is the mass of the hydrogen atom. The corresponding rate of the
kinetic energy injection needed to support the expansion is $\dot{M}_{\rm neb}v^2/2$.
The total power needed to support the nebula expansion is therefore
\begin{equation}
\dot{E}_{\rm exp}=M_{\rm neb}v \dot{v} + 2\pi R_{\rm neb}^2 n_{\rm H} m_{\rm H}v^3~.
\label{E.exp}
\end{equation}

Under our assumptions, $\dot{E}_{\rm rot}$ should exceed the sum $\dot{E}_{\rm
rad}+\dot{E}_{\rm exp}$. This condition results in a lower-bound on the moment of
inertia of the Crab pulsar,
\begin{equation}
I_{\rm Crab}>{\dot{E}_{\rm exp}+\dot{E}_{\rm rad}\over \Omega
\vert\dot{\Omega}\vert}~. \label{ICrab.Ebound}
\end{equation}
which can be rewritten as

\begin{eqnarray}
I_{\rm Crab,45}> 0.34\left({d\over {\rm 2~kpc}}\right)^2
 + 0.55\ {M_{\rm neb}\over M_\odot} + \nonumber \\ 
 \ 0.15 \left({R_{\rm neb}\over {\rm 1~pc}}\right)^2 {n_{\rm H}\over 0.2~{\rm
cm^{-3}}}~. \label{ICrab.Ebound.num}
\end{eqnarray}

We will use the canonical distance to the Crab nebula given by Davidson \& Fesen
(1985), $d=1.83$ kpc. The second term on the right-hand-side of Eq.\
(\ref{ICrab.Ebound.num}) is the most important one - alas, it is also the most
uncertain. Davidson \& Fesen (1985) give the estimate $2-3~M_\odot$, while
MacAlpine \& Uomoto (1991) evaluate this mass as  $1-2~M_\odot$. The most recent
estimate was obtained by Fesen et al. (1997), who get  much higher value $4.6\pm
1.8~M_\odot$. 

In what follows, we will consider two estimates of $M_{\rm neb}$. The lower
one, $2~M_{\odot}$, is a mean of the evaluations given by Davidson \& Fesen
(1985) and MacAlpine \& Uomoto (1991). The higher one, $4.6~M_{\odot}$, 
uses the "central value" of the mass calculated by Fesen et al. (1997).

Most recent  estimates of $n_{\rm H}$ are based on the evaluation of the hydrogen
column density in the direction of Crab. Such a method yields $n_{\rm H}\simeq
0.5~{\rm cm}^{-3}$ (Sollerman et al. 2000, Willingale et al. 2001). However, this
might be a severe overestimate of $n_{\rm H}$, because the Crab radiation goes through
the Galaxy disk on his way to the detector. In view of this, we will prefer to use a
conservative value $n_{\rm H}=0.2~{\rm cm}^{-3}$ quoted by Manchester \& Taylor
(1977). 

Using the above mentioned values of the quantities entering the
right-hand-side of Eq.\ (\ref{ICrab.Ebound.num}), we get two lower bounds
for the moment of inertia $I_{\rm Crab}$, corresponding to two different estimations 
of the Crab nebula mass:  
$I_{\rm Crab,45}>1.61$ for $M_{\rm neb}=2~M_{\odot}$, and 
$I_{\rm Crab,45}>3.04$ for $M_{\rm neb}=4.6~M_{\odot}$.

\section{The $I(M,R)$ correlation}
In their seminal paper, Ravenhall \& Pethick (1994) studied
the properties of the ratio $I/MR_\infty^2$, where $R_\infty$
stands for an apparent (radiation) stellar radius,
 determined  in  the analyzes of thermal radiation
from the close by neutron stars,
\begin{equation}
R_\infty = {R\over \sqrt{1-2GM/Rc^2}}~,
\label{Rinf.def}
\end{equation}
where $R$ is the coordinate radius of the space-time metric. In particular,
they found that for an intermediate range of $(M/M_\odot)({\rm km}/R)
\sim 0.05-0.15$ the ratio
$I/MR_\infty^2$ stays nearly constant, and depends rather weakly on the
dense matter EOS. This finding resulted in their approximate formula
for $I$.

\begin{figure}[h]
\centering
\resizebox{3.5in}{!}{\includegraphics{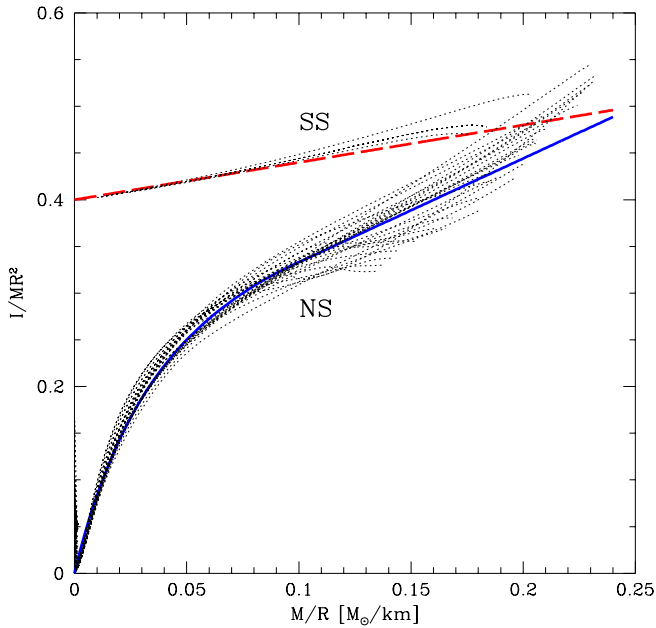}}
\caption{The ratio $I/M R^2$ versus compactness parameter
$M/R$ for selected set of EOSs of dense matter. Branches
NS and SS correspond to neutron stars and strange stars.
Thick lines were obtained using the
 analytical fitting formulae (solid - NS, dashed - SS),
described in
the text. Both thick lines are terminated at the $M/R$ upper-bound
for the EOS respecting condition $v_{\rm sound}\le c$, which is
equal $0.24$.
}
\label{fig:IMR2}
\end{figure}

In  our quest for a universal, i.e., independent of the specific EOS approximate
relation between $I$, $M$ and $R$, we relied on a large set of more than thirty
EOSs of dense matter.
 This set includes various models of dense matter, which were
obtained under different hypotheses on the composition of the matter at
$\rho>2\rho_0$, where $\rho_0 = 2.7\cdot 10^{14}~{\rm g~cm^{-3}}$ is the normal
nuclear density. Considered compositions of the super-dense matter in neutron stars
(NS) include: nucleons and leptons (EOSs of Baldo et al. 1997, Bombaci 1995,
Balberg \& Gal 1997; Balberg et al. 1999;
Bethe \& Johnson 1974, Pandharipande 1971,
Pandharipande \& Ravenhall 1989, Douchin \& Haensel 2001, Wiringa et al. 1988,
Walecka 1974, Haensel \& Pr{\'o}szy{\'n}ski 1980, Akmal et al. 1998);
 nucleons, hyperons, and leptons (EOSs of Glendenning 1985, 1997, Balberg
 \& Gal 1997, Weber et al. 1991, Balberg et al. 1999); deconfined quark
matter mixed with baryonic matter (EOSs of Glendenning 1997); pion and
kaon condensates coexisting with baryonic
matter,  in a pure or a mixed-phase configuration
(EOSs of Glendenning 1997, Kubis 2001). We considered also three models of
self-bound absolutely stable strange quark matter, forming hypothetical strange stars
(SS; EOS of Haensel et al. 1986).  After studying various types
of correlations between $I$, $M$ and $R$, we
found, in accordance with Lattimer \& Prakash (2001),  that it is the ratio $I/MR^2$
which has the most universal dependence on the compactness $M/R$.  Our result is seen
in Fig. \ref{fig:IMR2}.

In what follows,
we will use, following Ravenhall \& Pethick (1994), the
 dimensionless compactness parameter $x\equiv M/R\cdot
 {\rm km}/M_\odot$. The dependence of $I/MR^2$ on $x$ is well
 approximated, up to $M_{\rm max}$, by a function $a(x)$, different
 for neutron stars (NS) and strange stars (SS),
\begin{math}
\\
\\
a_{\rm NS}(x)=
\left\{
\begin{array}{ll}
x/( 0.1 +2x),\       & {\rm if\ } x\le\ 0.1\\
{2\over 9}(1+5x),\      & {\rm if\ } x>\ 0.1  \\
\end{array}
\right .
\label{a.NS}
\end{math}
\begin{equation}
a_{\rm SS}(x)={2\over 5}(1+x)~.
\label{a.SS}
\end{equation}

The difference in the $x$-dependence of $I/MR^2$ between
 neutron stars and strange stars
stems from the qualitative difference in their density profiles. The
density profile within a moderate mass strange star is  very flat.
On the contrary, density falls by many orders in the outer layer of
a neutron star. Consequently, at a same $x\le 0.2$,
$(I/MR^2)_{\rm SS}>(I/MR^2)_{\rm NS}$
(the case of $x>0.2$ will be discussed later in this section).
As we see in Fig. \ref{fig:IMR2}, the precision of the
formulae given in Eqs.\ (\ref{a.NS}) is
typically better than 10\%. The formula for neutron stars
breaks down only at very
low $M$, below $0.2~M_\odot$. Such low masses do not
seem relevant for neutron stars observed as pulsars.
\footnote[1]{Approximate formula for very low values of $M/R$ corresponding to the
vicinity of the $M_{\rm min}\simeq 0.1~M_\odot$ is given in
the footnote on p. 850 of Ravenhall \& Pethick (1994).(We believe, that
we have found a couple of misprints in this footnote: the equation for very
small $x=M/R$ should be rewritten as $I\simeq 7.5\cdot M^2 R$, and for greater $x$,
$I/MR^2\Lambda(R)\simeq 0.23 - 0.07\cdot(1-10x)^2\cdot \sqrt{x}$)}
 The  formula for $a_{\rm SS}$ looses its validity only for very
 low-mass ($M<0.1~M_\odot$) strange stars with crust.
 On  the contrary, the formula for $I$  becomes {\it exact} for
 low-mass {\it bare} strange stars, because they are Newtonian and
 their density is to a very good approximation constant, which
 corresponds to $a_{\rm SS}\simeq 2/5$.

The overall upper-bound on $x$ is reached at $M_{\rm max}$ for a
pure causality-limit EOS $P=c^2(\rho-\rho_{\rm s})$, and does
not depend on $\rho_{\rm s}$:
$x^{\rm CL}_{\rm max}=0.240$ (see Lindblom 1984, Haensel et al. 1999).
Within General Relativity, static compact star models  with
EOS  satisfying $v_{\rm sound}\le c$ have $x\le x^{\rm CL}_{\rm max}$.
We obtain $a_{\rm NS}(x^{\rm CL}_{\rm max})=0.489$ and
 $a_{\rm SS}(x^{\rm CL}_{\rm max})=0.496$, so that in the $x=x^{\rm CL}_{\rm max}$
limit $a_{\rm SS}\simeq a_{\rm NS}$ within about 1\%.
\section{Estimate of $I_{\rm Crab}$ and constraints in the $M-R$ plane}

\begin{figure}[h]
\centering
\resizebox{3.5in}{!}{\includegraphics{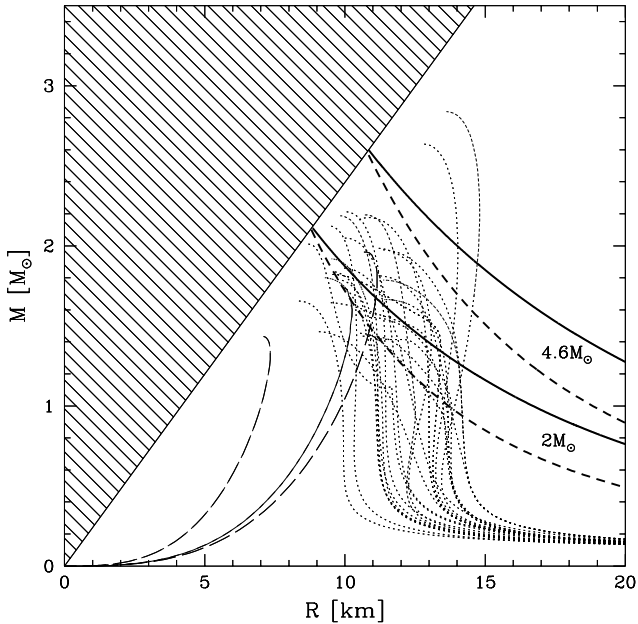}}

\caption{ The mass-radius curves for a large sample of EOS (thin dotted
curves - neutron stars, thin dashed curves - strange stars) and
constraints resulting from the estimates of $I$ of the Crab
pulsar (thick solid line - neutron stars, thick dashed line - strange
stars), obtained in the text. Hatched area is excluded by the
General Relativity combined with $v_{\rm sound}\le c$.}

\label{fig:CrabMR}
\end{figure}

{ 
In Sect.\ 2 we obtained two lower bounds for the Crab pulsar 
moment of inertia $I_{\rm Crab}$, which depend on the different evaluations of
the expanding shell mass: 
$I_{\rm Crab,45}>1.61$ for $M_{\rm neb}=2~M_{\odot}$ (conservative
estimate), and 
$I_{\rm Crab,45}>3.04$ for $M_{\rm neb}=4.6~M_{\odot}$ (newest estimate).
}

These bounds on $I_{\rm Crab}$, combined with our fitting formulae,
derived in Sect.\ 3, imply constraints on the mass and radius of 
the Crab pulsar (for details, see Fig.\ \ref{fig:CrabMR}).
Neutron star configurations below the thick solid lines are ruled
out. The weaker, conservative constraint eliminates a number of soft 
and moderately stiff EOS. 

{ For those EOS which are acceptable, this 
constraint gives the range of possible mass and radius of the Crab 
pulsar, $M_{\rm Crab}>1.2~M_\odot$ and $R_{\rm Crab}=10-14$~km.
The second ($M_{\rm neb}=4.6~M_{\odot}$) constraint is much more 
dramatic. Only a very stiff EOSs are accepted, because $M_{\rm
Crab}=1.9~M_{\odot}$ and $R_{\rm Crab}=14-15~~{\rm km}$.

For the strange-star model we obtain strong constraints 
(thick dashed line in Fig. 2) even for the conservative estimate 
$M_{\rm neb}=2~M_{\odot}$: $M_{\rm Crab}^{\rm (SS)} > 1.5\ M_{\odot},
R_{\rm Crab}^{\rm (SS)}=10-11\ {\rm km}$.
Putting $M_{\rm neb}=4.6~M_{\odot}$ excludes definitely the 
strange-star model.
}
Note, however, that the strange-star constraint is likely to be 
an academic one anyway because the glitches in the timing of the Crab pulsar 
are widely believed to rule-out the strange-star model 
(Alpar 1987, nevertheless see Glendenning \& Weber 1992).

The formula expressing $I$ in terms of $M$ and $R$ may be
useful for putting constraints, resulting for pulsar
observations,  on the neutron-star models. Consider a pulsar-powered
supernova remnant of the Crab type. If the {\it measured}
total power needed
to sustain the radiation of the system and the acceleration of the
nebula expansion is $\dot{E}_{\rm tot}$, this implies the lower
bound on the pulsar moment of inertia, $I>\dot{E}_{\rm tot}/
(\Omega\vert\dot{\Omega}\vert)$, where $\Omega$ and $\dot{\Omega}$ are {\it
measured} pulsar angular frequency and its time derivative. Reliable
estimates of $\dot{E}_{\rm tot}$, combined with ``empirical formula''
for $I$, would then imply constraints on the neutron-star models in
the $M-R$ plane.
Bounds for the mass and radius of the Crab pulsar obtained for
$M_{\rm neb}=4.6~M_{\odot}$ are very strong and should therefore be
considered with caution. More advanced models of the structure and 
evolution of the Crab nebula deserve to be developed. This problem
will be addressed in a subsequent paper.
\begin{acknowledgements}
We would like to thank Prof. B. Paczy{\' n}ski for helpful remarks.
This research
was partially supported by the KBN grant No. 5P03D.020.20.
\end{acknowledgements}
\section*{Appendix: Correlation between $I_{\rm max}$
and $M_{\rm max}$, $R_{M_{\rm max}}$}
The fact that $a_{\rm NS}$  and $a_{\rm SS}$ converge at the highest values of
compactness parameter suggests that there might exist a universal  correlation between
the maximum value of $I$, $I_{\rm max}$, and $M_{\rm max}$ and $R_{M_{\rm max}}$,
valid for both neutron stars and strange stars. Such a correlation has been pointed
out by Haensel (1990) for a set of nineteen EOS developed in 1970s and 1980s:
 $I_{\rm max,45} = 0.98\cdot (M_{\rm max}/M_\odot)
(R_{M_{\rm max}}/{\rm 10~km})^2$.
For the EOS set considered by Haensel (1990) this ``empirical formula''
reproduced values of $I_{\rm max}$ with an overall accuracy $\chi^2 = 0.03$.
\footnote[2]{In the present paper the value of $\chi^2$ is actually scaled by the
variance, i.e., it is equal to the standard value of $\chi^2$ divided by $\sigma^2$.
Such a treatment is convenient in our case, where  we have to compare the qualities of
different fits.}
Within our sample of numerical results, obtained for selected thirty EOSs,
we found correlation  $I_{\rm max}\simeq
 {\cal C} (M_{\rm max}/{\rm M}_\odot)
 (R_{M_{\rm max}}/10~{\rm km})^2$,
 the same as in (Haensel 1990), but  with a slightly lower numerical
factor ${\cal C}= 0.97$;
 the overall fit is characterized by $\chi^2=0.05$. The value of $\chi^2$
can be significantly reduced if one takes into account an additional correlation
 between the ratio $I_{\rm max}/(M_{\rm max}R_{M_{\rm max}}^2)$
and the maximum compactness parameter $x_{\rm max}$ reached at $M_{\rm max}$. The
improved empirical formula yields:
\begin{equation}
I_{\rm max,45}\simeq (-0.368 + 7.122\cdot x_{\rm max})
 \left({M_{\rm max}\over {\rm M}_\odot}\right)
\left({R_{M_{\rm max}}\over {\rm 10~km}}\right)^2~. \label{Imax.eq2}
\end{equation}
It is  valid  both for neutron stars and strange stars, with the overall quality of
the fit $\chi^2=0.02$. The coefficient ${\cal C}=0.97$  in the
simplest empirical formula corresponds to an
``average value of $x_{\rm max}$'', $\bar{x}_{\rm max}=0.19$. Some general statements
concerning different classes of the EOS can be made. The ``empirical formula'', Eq.
(\ref{Imax.eq2}),  under-estimates $I_{\rm max}$ for the superluminal EOSs. On the
contrary, it overestimates $I_{\rm max}$ for the EOSs with high-density softening due
to the appearance of hyperons. Finally, for the ``minimal composition EOSs'' (nucleons
and leptons) as well as for the strange matter EOSs, our  formula reproduces $I_{\rm
max}$ within better than $5\%$.

A different, very precise formula for $I_{\rm max}$ of strange stars (bare or with
crust) results from ``scaling properties'' (Haensel et al. 1986) of the configurations
with $M_{\rm max}$ and $I_{\rm max}$. For the simplest model of strange quark matter
composed of massless, non-interacting quarks, scaling with respect to the value of the
bag constant implies {\it exact} relation $I_{\rm max,45}=0.94
 \left({M_{\rm max}/M_\odot}\right)
\left({R_{M_{\rm max}}/{\rm 10~km}}\right)^2$. After inclusion of the QCD interaction
and finite strange-quark mass the above equation is no longer exact.
 For $m_{\rm s}c^2=(150-200)~$MeV, the numerical prefactor to be put
in the expression for $I_{\rm max,45}$
 is  0.97, and the formula is precise within a few percent.

 One might argue that the validity of a simplest  "empirical formulae"
for $I_{\rm max}$  is a  consequence of the fact
that {\it realistic}
EOSs are well approximated by the polytropes. For a polytropic EOS
$P=K n_{\rm b}^\Gamma$ (where $n_{\rm b}$ is baryon density of matter)
with a fixed $\Gamma$, the  relation $I_{\rm max,45}={\cal C}
\cdot (M/M_\odot)(R_{M_{\rm max}}/10~{\rm km})^2$  is {\it exact},
with ${\cal C}$ independent of $K$. We get ${\cal C}_{\Gamma=2}
=0.83$ and ${\cal C}_{\Gamma=3}=1.04$, so that the "best fit" value
for realistic EOSs lies roughly in the middle of the $\Gamma=3$
and $\Gamma=2$ values. However, the realistic EOSs
 at super-nuclear densities
{\it are not polytropes}. One can speak only about a local, density dependent
$\Gamma(n_{\rm b})$.
 For example, in the
case of dense matter with hyperons the EOS softens considerably
(and in a discontinuous way) at the thresholds for the
appearance of a new hyperon, with $\Gamma$ dropping below one
(Balberg \& Gal 1997). Clearly, the validity of the empirical
formula for $I_{\rm max}$ is not due to the polytropic character
of the EOS of dense matter.

\end{document}